\documentclass [twocolumn,
amsymb,amsmath,nobibnotes,superscriptaddress,aps,prb]{revtex4}
 \usepackage {bm}
 \usepackage{graphicx}
 \usepackage{amsmath}

\begin{document}
 \title{Giant mesoscopic spin Hall effect on surface of topological insulator}
 \author{Jin-Hua Gao}
 \affiliation{Department of Physics, and Center of Theoretical and Computational Physics, The University of Hong Kong,
 Hong Kong, China}
\author{Jie Yuan}
 \affiliation{Department of Physics, and Center of Theoretical and Computational Physics, The University of Hong Kong,
 Hong Kong, China}

 \author{Wei-Qiang Chen}
 \affiliation{Department of Physics, and Center of Theoretical and Computational Physics, The University of Hong Kong,
 Hong Kong, China}
\author{Yi Zhou}
 \affiliation{Department of Physics, Zhejiang University, Hangzhou,  China}
 \author{Fu-Chun Zhang}
 \affiliation{Department of Physics, and Center of Theoretical and Computational Physics, The University of Hong Kong,
 Hong Kong, China}
\affiliation{Department of Physics, Zhejiang University, Hangzhou,
China}
 \begin{abstract}
 We study mesoscopic spin Hall effect on the surface of topological insulator
 with a step-function potential.
 The giant spin polarization induced by a transverse electric current
 is derived analytically by using McMillan method
 in the ballistic transport limit, which oscillates across
 the potential boundary with no confinement from the potential barrier
 due to the Klein paradox, and should be observable in spin
 resolved scanning tunneling microscope.
  \end{abstract}
  \maketitle
Topological insulator (TI) with time reversal invariance has
recently been proposed theoretically and observed in
experiments\cite{kane1,sczhang1,sczhangrev,kanerev, moore, hasan1,
hasan2,xue,fang}. In three spatial dimensional (3D) TI, electronic
structure is characterized by a bulk gap and a gapless surface mode
described by odd number of branches of Dirac particles, which is
protected by time reversal symmetry. The surface states are helical,
where spins are locked with momentum. The gapless Dirac dispersion
mode of the surface states has  been confirmed in angle resolved
photoemission spectroscope, and the predicted spin-momentum
correlation has also been reported in
experiments\cite{hasan1,hasan2}. Because of the strong correlation
between the spin and momentum, the surface states of 3D TI may be a
potentially idea system to study spintronics, where the electron's
spin degree is used to manipulate and to control mesoscopic
electronic devices. It is interesting to note that the recent
progress in TI is closely related to the development of the spin
Hall effect (SHE) in the past several years, a sub-topic in
spintronics.

The SHE refers to a boundary (surface or edge) spin polarization when
an electric current is flowing through the system.
There have been extensive studies on the SHE in the conventional semiconductors
or metals with spin-orbit coupling,
 both in experiment\cite{kato, sih, wunderlich, valenzuela}
and in theory\cite{hirsch, sfzhang, engel, tse, murakami, sinova}.
 The SHE is often classified into
 "extrinsic" (impurity driven) or "intrinsic" (band structure driven).
Since arbitrarily weak disorder destroys the intrinsic SHE in 2D
infinite system with linear spin-orbit
coupling\cite{murakami2,engel2}, there have been considerable
interests on the mesoscopic systems in the ballistic limit, where
the disorder may be ignored
 \cite{nikolic, zyuzin, usaj, jyao, yxing, bokes}.
In the ballistic limit, the electric field is absent inside the
system, and the spin polarization is resulted from spin precession
around the lateral confined potential. The SHE of 2D semiconductor
system in the ballistic limit has been studied
theoretically\cite{zyuzin,usaj,yxing}. While the ballistic spin
accumulation is predicted near the potential barrier, the effect has
not been observed in experiments for the weakness of the effect in
realistic semiconductors or for the difficulties to detect the
spatial distribution of the spin polarization in the sandwiched
interface. The surface states of the TI represent a different type
of 2D system where the spin-orbit coupling is strong, and the
surface state can be probed directly by scanning tunneling
microscope (STM). This may provide a new route in study of the
SHE\cite{silvestrov, sqshen} and spintronics in general.

In this work, we report theoretical prediction of a giant SHE on a
surface of 3D TI with a step-function potential in the ballistic
limit as schematically illustrated in Fig. 1. By using McMillan
Green function method, we derive analytic expressions for the
electric current-induced spin polarization, which is oscillating
across the potential boundary, and is not confined by the potential
due to the Klein paradox. For a typical TI, the amplitude of the
local spin polarization is estimated to be as large as  $20\%$ at
the Fermi level near the boundary, which is much larger than the SHE
in a typical 2D semiconductor system, and should be observable in
spin resolved STM experiments.

\begin{figure}
\centering
\includegraphics[width=7cm]{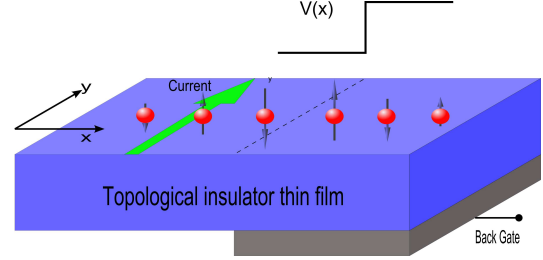}
\caption{(Color online) Schematic illustration of the proposed
ballistic spin Hall effect on the surface ($x-y$ plane) of a
topological insulator. A back gate of voltage $V_2$ is applied at
the right side with $x>0$. Electric current flows along
$y$-direction, resulting in spin polarization oscillation along
$x$-direction across the barrier.} \label{model}
\end  {figure}

 We consider surface states of TI, described by an effective Hamiltonian in
 the x-y plane,
  \begin{equation}\label{hamiltonian}
   H= v_F(\bm{p} \times \bm{\sigma} )\cdot \bm{\hat{z}}  + V(x),
  \end{equation}
  where $\bm{p}$ is the electron momentum,
  $\bm{\sigma}$ are the Pauli matrices, and
  $v_F$ is the Fermi velocity.  The system is translational invariant along the y-axis, and
  has a step-function potential at $x=0$, which separates two
  regions along the x-axis: region 1 at $x<0$ and region 2 at
  $x>0$, as illustrated in Fig. 1,
  \[ V(x) = \left\{ \begin{array}{ll}
         V_1=0 & \mbox{if $ \ x < 0$};\\
         V_2 & \mbox{if $ \ x > 0$}.\end{array} \right. \]
 where $V_2$ is a constant.  A voltage of $V_y$ is applied across the surface to induce an electric current along the $y$-direction.
 We consider the ballistic limit, where the electric field inside the surface is zero. We will first construct the
 retarded Green's function by using the scattering wavefunction, a method introduced by McMillan to study superconducting state. From the obtained Green functions,
 we calculate the local spin density in the presence of the electric current to show the profound SHE. The experimental consequences and comparisons with the SHE in conventional 2D semiconductors will be discussed.

 The scattering wave functions can be constructed based on the eigen functions of the Dirac particle in Hamiltonian (1) in the two separate spatial regions.  The
 eigen wavefucntions in region $\alpha$ corresponding to the energy
 $\epsilon$ and y-component momentum $k_y$ are given by
\begin{equation}\label{eigf 1}
 \varphi^{\pm}_{\alpha}(x,y)= e^{ i(\pm k_{\alpha x} \cdot x + k_{ y}  \cdot y )}
 \left(\begin{array}{cccc}
1 \\ i \hbar v_F \frac{\pm k_{\alpha x} + i k_{y}}{\epsilon - V_\alpha} \\
\end{array}\right)
\end{equation}
where $\hbar v_F k_{\alpha x} = \sqrt{(\epsilon - V_\alpha) ^2 -
(\hbar v_F k_{y})^2}$.
By adjusting the gate
potential $V_\alpha$ relative to the Fermi energy $E_F$, the Dirac
fermion carriers in region $\alpha$ can be tuned into electron-like
 ($n$-type, $\textrm{E}_F > V_\alpha$) or hole-like ($p$-type,  $\textrm{E}_F < V_\alpha$
 ). Therefore the system may be viewed as  $n-n$ or $n-p$ types of junction.

 The right ($R$) and left ($L$) moving scattering wavefunctions can then be found by using the standard transfer matrix
method. For a $n-n$ junction, we have
\begin{eqnarray}\label{scatteringnn}
\phi^{nn}_R(x,y) &=& \left\{
\begin{array}{llll}
             \varphi^{+}_1 (x,y) + r^{nn}_R \varphi^{-}_1 (x,y)&  \mbox{if $x < 0$}   \\
            t^{nn}_R \varphi^{+}_2 (x,y) &  \mbox{if $x > 0.$}    \\

        \end{array}
\right.  \notag \\
\phi^{nn}_L(x,y)&=& \left\{
\begin{array}{llll}
              t^{nn}_L \varphi^{-}_1 (x,y)&  \mbox{if $x < 0$}   \\
              \varphi^{-}_2 (x,y) + r^{nn}_L \varphi^{+}_2 (x,y) &  \mbox{if $x > 0.$}    \\

        \end{array}
\right.
 \end{eqnarray}

\noindent where $t^{nn}_{R/L}$ and $r^{nn}_{R/L}$ are the
transmission and reflection coefficients respectively, which are
related by
\begin{align}
\left( \begin{array}{llll}
         t^{nn}_R \\ 0
         \end{array}
  \right) = T
  \left( \begin{array}{llll}
         1 \\ r^{nn}_R
         \end{array}
  \right), \,\,
  \left( \begin{array}{llll}
        r^{nn}_L \\ 1
         \end{array}
  \right) = T
  \left( \begin{array}{llll}
         0 \\ t^{nn}_L
         \end{array}
  \right) \nonumber
\end{align}

\noindent where $T$ is the transfer matrix,
\begin{equation}
T = \frac{\epsilon - V_2}{2 k_{2x}} \left( \begin{array}{llll}
          \frac{k_{2x} - i k_y}{\epsilon - V_2} + \frac{k_{1x} + i k_y}{\epsilon - V_1} &
          \frac{k_{2x} - i k_y}{\epsilon - V_2} - \frac{k_{1x} - i k_y}{\epsilon - V_1} \\

          \frac{k_{2x} + i k_y}{\epsilon - V_2} - \frac{k_{1x} + i k_y}{\epsilon - V_1} &
          \frac{k_{2x} + i k_y}{\epsilon - V_2} + \frac{k_{1x} - i k_y}{\epsilon - V_1} \\
       \end{array}
\right) \nonumber
\end{equation}
The scattering wavefunctions for a $n-p$ junction have similar
form with that for the $n-n$ junction, except that
$\phi^{+(-)}_{2}$ are replaced by $\phi^{-(+)}_{2}$ in the region
of $x>0$ for the group velocity of a hole is opposite to that of
an electron and that all the superindices of $nn$ are replaced by $np$.

We note that if $k_x$ is complex, the evanescent wave appears.
In this case, considering the asymptotic behavior of the
evanescent wave, the scattering wave function for both $n-n$ and $n-p$
junction will have the form given in Eqn. \eqref{scatteringnn}.

We are interested
in the transverse effect of the charge and spin density profiles as
an electric voltage is applied along the y-axis. To this end we construct
the retarded Green's function, which satisfies the
equation,
\begin{equation}
(\epsilon - H) G^r(x,x';\epsilon, k_y)= \delta (x - x') I
\end{equation}
where $I$ is a 2 by 2 identity matrix. The solution for $G^r$ is a
direct product \cite{mcmillan} of the scattering wavefunctions
$\phi_{R/L}$ and the transposal wavefunctions $\hat{\phi}^t_{L/R}$,
\begin{equation}\label{gf}
G^r(x,x'; \epsilon, k_y) = \left\{ \begin{array}{llll}
                           c^{<}\phi_L(x,y) \hat{\phi}^t_R(x',y) & \mbox{if $x < x'$}\\
                           c^{>}\phi_R(x,y) \hat{\phi}^t_L(x',y) & \mbox{if $x > x'$}\\
                    \end{array}
                    \right.
\end{equation}
where $c^{<}$ and $c^{>}$ are the coefficients, which can be
determined from Eqn. (4). Here, $\hat{\phi}_{L/R}$ has the same
form as $\phi_{L/R}$, expect the replacement of the factor
$e^{ik_y \cdot y}$ by $e^{-ik_y \cdot y}$.

The local  spin density of states for a giving $k_y$ ($\mathbf{S}$) and the local charge density of states $\rho$
at energy $\epsilon$ can be found easily from $G^{r}$,
 \begin{eqnarray}
 \bm{S}(x; \epsilon, k_y) &=& - \frac{\hbar}{2\pi} \textrm{Im}\textrm{Tr} [G^r(x,x; \epsilon, k_y)
 \bm{\sigma}], \nonumber\\
\rho (x,\epsilon) &=& - \frac{1}{\pi} \sum_{k_y}\textrm{Im} \textrm{Tr} G^r(x,x; \epsilon, k_y),\\
\end{eqnarray}
where the sum in $\rho$ is over all the possible values of $k_y$, and the contributions from the evanescent waves are also included.

In the present case, the Green's function, hence the local charge
and spin density of states can all be solved analytically.  Here
we shall focus on the spin z-component, which is most interesting
and given by
\begin{eqnarray}\label{sz}
 &&\textrm{Tr}[G^r(x; \epsilon, k_y)\sigma_z] = F(\epsilon, k_y)[e^{2is_2k_{2x}x} \Theta(x)  + e^{-2ik_{1x}x}\Theta(-x) ]
  \nonumber \\
 &&F(\epsilon,k_y) = \frac{\frac{k_y}{\epsilon - V_1} - \frac{k_y}{\epsilon - V_2}}{1 + \frac{(\hbar v_F)^2 (s_2k_{1x} k_{2x} - k^2_y)}{(\epsilon - V_1)(\epsilon - V_2)}},
 \end{eqnarray}
where $\Theta(x)$ is Heaviside function and $s_2 = sign(\epsilon -
  V_2)$.  The charge density of states is directly related to the trace of $G^r$,
\begin{widetext}
\begin{equation}
\textrm{Tr}[G^r(x;\epsilon, k_y)] = \frac{\Theta(x)(\epsilon -
V_2)}{i  (\hbar v_F)^2 s_2 k_{2x}}[1 + F(x;\epsilon, k_y)
\frac{(\hbar v_F)^2 k_y}{\epsilon - V_2}  e^{2i  s_2 k_{2x}x}] +
\frac{\Theta(-x)(\epsilon - V_1)}{i  (\hbar v_F)^2 k_{1x}} [1 -
F(\epsilon,k_y) \frac{(\hbar v_F)^2 k_y}{\epsilon - V_1}
e^{-2ik_{1x}x}]
\end{equation}
\end{widetext}

\begin {figure}
\centering
\includegraphics[width=8.5cm]{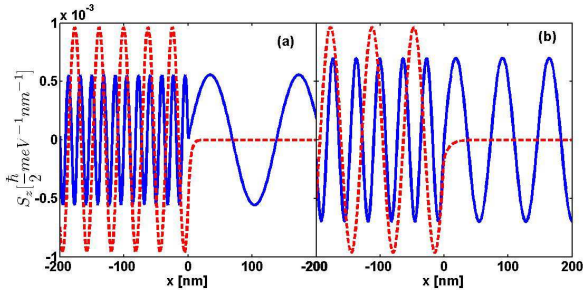}
\caption{(Color online) Spatial distribution of local spin density
$S_z$ for a single Dirac fermion in Hamiltonian (1) with
 $v_F = 5 \times 10^5 \ m/s$.
(a) n-n junction,  $V_2 = 40 \ \textrm{meV}$, and $\epsilon = 60 \
\textrm{meV}$.  $\theta = 63^\circ$ (blue solid line) and $\theta =
18^\circ$ (red dashed line). (b) n-p junction, $V_2 = 60 \
\textrm{meV}$ and $\epsilon = 35 \ \textrm{meV}$.  $\theta =
63^\circ$ (blue solid line) and $\theta = 36^\circ$ (red dashed
line).   $\theta$ is an incident angle, defined by $e^{i\theta} =
\frac{k_x + ik_y}{|\epsilon - V|}$.} \label{szall}
\end  {figure}

The local spin density of states for a giving $k_y$ in the $n-n$
and $n-p$ junctions are plotted in Fig.\ref{szall}.  One important
feature is the non-confinement of the Dirac particle with the
higher barrier potential ($V_2 > \epsilon$)  due to  the Klein
paradox. The local spin density  strongly depends on the incident
angle of the electron, as we can see from Fig. \ref{szall}. The
evanescent wave appears if $|\epsilon - V_1| > \hbar v_F k_y >
|\epsilon - V_2|$, and there is a critical incident angle for the
condition of the evanescent wave. These features are typical
characteristics of Dirac fermion, essentially the same as in the
graphene.

\begin {figure} \centering
\includegraphics[width=8.5cm]{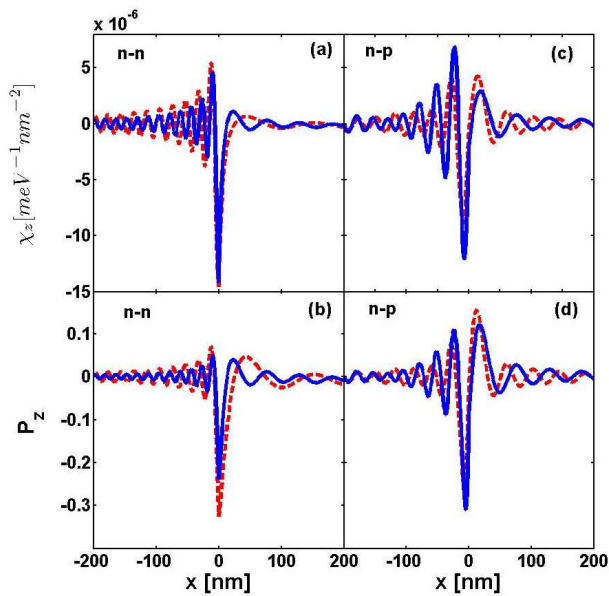}
\caption{(Color online)  Spatial profiles of spin polarization
density  $\chi_z(x;E_F)$ (top panels), and spin polarization
$P_z(x; E_F)$ (bottom panels) on the surface of TI, described by
Eq. (1) with $v_F=5 \times 10^5 m/s$. Left panels (a) and (b) are
for $n-n$ junction: $V_1 = 0$, $V_2 = 40\ \textrm{meV}$, and the
Fermi energy $E_F = 50 \ \textrm{meV}$ (red dashed line) and $E_F
= 60 \ \textrm{meV}$ (blue solid line).  Right panels (c) and (d)
are for $n-p$ junction: $V_1 = 0$, $V_2 = 60\ \textrm{meV}$, and
$E_F=  35 \ \textrm{meV} $ (red dashed lines) and $E_F =  40 \
\textrm{meV}$ (blue solid lines). $P_z =1$ corresponds to
completely polarized spins at $E_F$. } \label{nn}
\end  {figure}

We now  calculate the spin polarization along x-direction near the
potential boundary $x=0$ induced by an electric current along the
y-direction at zero temperature.  We consider a voltage of $V_y/2$
at the one edge and $-V_y/2$ at the other edge of the system along
the y-axis, and consider the ballistic transport limit. The effect
of the voltage at the two edges is to induce an imbalance of the
occupied states between $k_y
>0$ and $k_y <0$. The states with $k_y>0$ are
occupied at energies below $E_F + V_y/2$, and the states with
$k_y<0$ are occupied at energies below $E_F - V_y/2$, with $E_F$ the
Fermi energy at $V_y=0$.  By the time reversal symmetry, the local
spin polarizations contributed from $k_y>0$ and from $k_y<0$ with
the same energy cancel to each other.  For small value of $V_y$, we
thus obtain the current-induced net spin density
profile\cite{zyuzin,usaj,datta}
\begin{equation}\label{tsz}
S^{in}_z(x;E_F)\simeq |eV_y| \sum_{k_y > 0} S_z(x;E_F,k_y)
\end{equation}
To further analyze the current-induced spin polarization, we
define local spin susceptibility $\chi_z(x; E_F)$ and local spin
polarization $P_z(x;E_F)$,
\begin{eqnarray}
\chi_z(x; E_F)&=& S^{in}_z(x; E_F) / \frac{\hbar}{2}eV_y \nonumber \\
P_z(x;E_F) &=& \chi_z(x; E_F) /\rho(x; E_F).
\end{eqnarray}
The local spin polarization $P_z$ is a dimensionless parameter to
measure the magnitude of the SHE. $P_z=1$ means the spins of the
electrons at a space point $x$ at the Fermi level are polarized. The
experimentally measured local spin density is obtained by
multiplying $P_z$ by the applied voltage and by the local density of
states. In Fig.\ref{nn} we plot $\chi_z$ and the $P_z$ in both $n-n$
and $n-p$ junctions.  The key features of the SHE in the system are
summarized below.  1).  There is a pronounced oscillation of spin
polarization near the potenial boundary $x=0$. The peak value of
$\chi_z$ is order of $10^{-6} \textrm{meV}^{-1}nm^{-2}$, and the
peak value of $P_z$ is about $20\%$, indicating the SHE here is
giant.  2). The induced spin polarization is found to be insensitive
to the Fermi energy on the TI surface. This may be understood
because the spin polarization is approximately inversely
proportional to $v_F$, and  $v_F$ is a constant for Dirac particles.
This is markedly different from the usual 2D system where $v_F$ is
proportional to $\sqrt{E_F}$, hence the spin polarization $~
1/\sqrt{E_F}$. 3). The oscillation period at the zero gate region is
inversely proportional to the Fermi wave vector, or $~1/k_F=\hbar
v_F/E_F$, typically tens of nm for $E_F$ at tens of meV, and the
period at the bottom gated region is proportional to $\hbar
v_F/|E_F-V_2|$, and can be larger. The spin polarization predicted
in our theory may be detected in spin resolved STM on mesoscopic
samples with over several micron mean free path. Different from the
2D electron gas in semiconductors formed in sandwiched interface,
which is difficult to use STM, the surface of TI can be directly
measured by STM. The bottom gated device allows to detect the spin
polarizations at both left and right regions. And we notice that the
back-gated TI thin film device has been realized in experiment
recently.\cite{lilu}

To further illustrate the giant SHE on the TI surface, we compare
our result with the theoretically estimated SHE in ballistic 2D
electron system (2DES) with Rashba spin-orbit coupling. Note that
the ballistic SHE in 2DES is relatively weak and has not been
observed in experiment. In the 2DES case, the Fermi velocity is
proportional to square root of the Fermi energy, so the spin
polarization is inversely proportional to the square root of the
Fermi energy\cite{zyuzin}. For a typical semiconductor such as
InGaAs/InAlAs heterostructure \cite{material}, we  have effective
mass $m^* = 0.05 m_e$, with $m_e$ the free electron mass, and the
Rashba spin-orbit coupling $\alpha_R \approx 10 \textrm{meV} \cdot
\textrm{nm}$ . The theoretical calculation\cite{zyuzin} indicates
that the peak value of $P_z \approx 2.4 \%$ if $E_F \approx 3.3
\textrm{meV}$, and $P_z \approx 0.4 \%$ for a more realistic Fermi
energy
 $E_F \approx 100 \textrm{meV}$ corresponding to the 2D electron density of
 $n_{2D} \approx 2.1 \times 10^{12}\textrm{cm}^{-2}$. Therefore, the
 local spin polarization we predicted for the TI surface is about 50 times
 larger than that in a typical 2D semiconductor with
 Rashba spin-orbit coupling.
 We also remark that the SHE we predicted in the TI is about 1000 times
  larger than  that estimated for HgTe quantum well where
 the spin-orbit coupling is induced by in-plane potential
  gradient.~\cite{yxing,material2}
We conclude that the surface of TI should be an excellent candidate
to observe the SHE in the ballistic limit.

In summary, we have theoretically examined the mesoscopic spin Hall
effect on a surface with a step function potential in 3-dimensional
topological insulator. By applying the McMillan Green's function
method, which is based on the scattering wave function method and
was previously used in study of superconductor junctions, we have
derived analytic expressions for the electric current-induced spin
polarizations on the surface (actually, this analytical method is
suitable for the ballistic SHE in various systems~\cite{ours}). In
the ballistic transport limit, a giant spin polarization oscillation
across the junction is induced by a transverse electric current. The
spin polarization is estimated as large as $20\%$, which is
insensitive to the Fermi level and is not confined by the potential
step due to the Klein paradox. Its magnitude is about two orders
larger than that in 2-dimensional electron gas with Rashba
spin-orbit coupling. The spatial oscillation period is order of
inverse of the Fermi wavevector. These features are markedly
distinguished from the 2-dimensional electron gas and should be
observable in experiments such as spin resolved scanning tunneling
microscope.

We acknowledge part of financial support from HKSAR RGC grant 701010
and CRF HKU 707010. YZ is partially supported by National Basic
Research Program of China (973 Program, No.2011CB605903), the
National Natural Science Foundation of China(Grant No.11074218) and
the Fundamental Research Funds for the Central Universities in
China.

 \end{document}